\documentstyle[aps,12pt]{revtex}
\begin{document}
\title{Past events never come back}
\author{X. de Hemptinne}
\address{Department of Chemistry, Catholic University of Leuven,\\ Celestijnenlaan 200 F, B-3001 Heverlee, Belgium\\
Address for correspondence: Duivenstraat 78, B-3052 Blanden, Belgium\\
e-mail: Xavier.deHemptinne@chem.kuleuven.ac.be}
\maketitle
\vskip 8mm
{\centerline {\bf ABSTRACT}}
\begin{abstract}
Time can be defined as the duration between events. It is irreversible. When used as a variable in quantifying the changing properties of matter, this irreversibility of time is incompatible with Newton's deterministic mechanics and with its quantum mechanical prolongation. Experimental evidence, involving rejection of the traditional isolation paradigm, points to a solution of this paradox. No system may be thought of as isolated. Irreversibility is imported from ``else\-where''. The popular objection of the followers of determinism, consisting in extrapolating local behaviour to the scale of the universe and stressing alleged contradictions generated by the non--isolation hypothesis, is intellectually unacceptable. It is unjustifiable to draw conclusions by extrapolating beyond the domain of scientific observation. 
\end{abstract}

\section{Introduction}

Time and space are fundamental concepts, which continue to resist all acceptable definitions. No discourse, by philosophers or scientists throughout the ages, has been able to force either notion into the strict intellectual construction required by human thinking. Both concepts lack the absolute frame of reference which is experienced in daily life and on which Newton built his classical mechanics of motion using time and space as the variables. Scientific interpretation imposes restrictions, but the picture inherited from mechanics, where the variables are interconnected by suitable differential equations, remains an unsatisfactory theoretical construction.

Newtonian mechanics has been revisited by Einstein with his theory of relativity, which gives the universe a curved non--Euclidean geometry. While this elegantly disposes of the absence of an absolute frame of reference for space, time remains a problem. Observers moving with respect to one another appear to live each with his or her own time scale. Minkowski unravelled the paradox by assigning properties that are mathematically connected and similar to that of space to the concept {\em time}.

While cosmologists continue to dispute mathematical models for the universe, it is clearly felt in daily life that time and space behave very differently indeed. Space concerns distances between objects while time is the field in which duration of events is measured. In contrast to our spatial environment, time has a direction. As it is said: {\em Time flies like an arrow}. Past and future are different and can never be made to coincide. Nature is by essence irreversible. In that context, Heracleitos, the ancient philosopher of Ephese, claimed $\Pi \acute\alpha \nu\tau\alpha \; ` \!\rho \widetilde{\epsilon\iota}$ all things flow; all things pass. Space does not have this property.

The present contribution concerns the problem of the irreversible evolution of phenomena at daily life level: its meaning, its mechanism and its origin. Although restricted in its philosophical ambition, the subject is extremely instructive as soon as we try to relate theoretical predictions to experimental facts. Although there has been continuous activity in this field since Boltzmann's attempts to rationalise dynamics using Newtonian mechanics, there has been an increase of interest in recent decades, spurred by the development of the mathematical theory of chaos.

In current literature, the word ``chaos'' has different meanings, usually not well distinguished from each other. This work aims also to clarify which is related to loss of memory of past events. 

In the local domain considered here, Newtonian mechanics is valid as has been proven over the years. Its possible quantum mechanical extension will be neglected in a first step. Newton's general equation for dynamics, relating the acceleration $a$ of any object with mass $m$ to a force $F$ is:
\begin{equation}
ma = F.
\end{equation}
According to Newtonian mechanics, the detailed evolution of space co--ordinates in the course of time (trajectory) depends on the initial conditions assumed to the motion, i.e. the spatial and velocity co--ordinates at the presumed initial instant ($t = 0$).

Without external intervention, forces are explicitly time invariant. Acceleration itself, as a second derivative of position with respect to time, is invariant under time reversal symmetry. It means that the artificial change of variable $t$ into $-t$ changes neither its value nor its sign so that this mathematical operation keeps the conclusions unchanged. Hence, according to the basic equation, Newtonian mechanics is perfectly time reversible.

In general, reversibility holds for any isolated system where the forces between the constituents have a zero sum. The laws of motion are indeed symmetrical with respect to inversion of the parameter {\em time}. Hence, no matter how intricate may be the particular trajectories of the system as time evolves, they preserve the memory of the initial conditions. This conclusion contradicts our general experience concerning the macroscopic property of nature: systems removed from their equilibrium state tend more or less quickly to lose the memory of their past history and spontaneously and irreversibly to reach their equilibrium state.

In taking isolated conditions as the basic hypothesis, Boltzmann was confronted with this incompatibility between the irreversible character of macroscopic dynamics and the time reversibility of Newtonian mechanics. To escape this contradiction, he assumed that the system would, by some unknown mechanism, reach and maintain what he called {\em molecular chaos} between its components. This chaos is a condition where no correlation whatever exists between individual particle motions. The mechanism for removing correlation of molecular motion was left unspecified but it was conjectured that the great number of collisions or interactions between the system's components and the complexity of the mechanics involved would be sufficient to account for it. In this, Boltzmann was violently attacked by his colleagues Zermelo and Loschmidt. Collisions, no matter how complex and numerous, are mechanical events responding perfectly to Newton's laws of mechanics which dictate their strict symmetry properties, in particular that with respect to time reversal.

Despite the early time controversies, Boltzmann's proposal remains today the basis for most fundamental investigations concerning the irreversible character of nature and the quest for theoretical predictions of its phenomenological consequences (transport coefficients). Clearly, the debate is not closed (Prigogine~{\em et al.}~1988, Lebowitz~1993) \nocite{prig:88,leb:93}.

The vast contemporary literature replaces Boltzmann's initial molecular chaos assumption by more detailed arguments derived from mathematical developments on ergodic theory, which address the concept {\em deterministic chaos} (at microscopic level) (Sinai~1976, Eeckmann {\em et al.}~1985, Ruelle~1989). \nocite{sinai:76,Eckmann:85,Ruelle:89}

The adjective {\em deterministic} points to evolutions where each state has a unique consequence. As such, it is opposed to the words {\em stochastic} or {\em random}. Newtonian mechanics for isolated systems is strictly deterministic at microscopic level. This property suggested to Laplace the dictum {\em ``given precise knowledge of the initial conditions, it should be possible to predict the future of the universe''}. 

{\em Chaos} is defined in the literature as the property of motion where long--term behaviour is unpredictable. It must be emphasised that, given perfect knowledge of the initial conditions, a deterministic dynamical system is perfectly predictable. In putting forward {\em deterministic chaos}, contemporary literature ascribes unpredictability to very sensitive and unstable dynamics coupled to uncertainty concerning initial conditions. Would therefore unpredictability and its consequence, irreversibility, be a logical inference of our personal lack of knowledge? This contradiction, amplified by the extreme positivistic attitude of some modern schools of mathematics, where formalism is preferred to experimental logic, generates a feeling of uneasiness in the scientific community, often hidden, but sometimes expressed formally (Dorfman~{\em et al.}~1995) \nocite{Dorfman:95}. 

Ambiguous semantics is the gateway to misunderstandings. Most controversies arise from unsettled disagreements in fundamental definitions. When focusing on time dependence, words like {\em irreversibility, isolation, equilibrium,} need further clarification.

The colloquial meaning of irreversibility implies the absence of spontaneous recurrence of particular conditions that would have been valid at some past instant. It will become obvious that this definition is far too weak. Our experience of nature suggests a stronger definition, where the word refers to complete loss of correlation or memory in going from past to future.

Daily experience teaches that perturbed macroscopic systems (consisting of many particles) tend to relax more or less rapidly until they reach an equilibrium state. The present article focuses on the time dependence of this evolution. For this to be discussed, an accurate definition of equilibrium is also required. It will soon become obvious that quantifying the equilibrium state of any macroscopic system is a vain exercise if the properties of the surroundings are ignored. The necessary intervention of the environment in relaxation dynamics and in the ultimate equilibrium conditions contradicts the generally accepted isolation paradigm implied by deterministic Newtonian mechanics at the microscopic (atomic or molecular) scale.

In the next section, strong experimental evidence for stochastic intervention of the surroundings will be highlighted on the basis of Joule's experiment. Section {\bf III} lists a number of objections raised against the non--isolation paradigm and refutes them. In section {\bf IV} we shall demonstrate that the dynamics involving stochastic exchange with the surroundings are not purely Newtonian. Thermodynamics is the appropriate tool for describing the interplay between systems removed from equilibrium and their environment. This will be generalised in section {\bf V} to the particular conditions valid out of equilibrium. Section {\bf VI} will examine the role of quantum mechanics.

\section{Relaxation scenario} 

Discussion of irreversible macroscopic dynamics is traditionally illustrated by the observation of the spontaneous expansion of a gas initially compressed into a fraction of what will become its final volume. The system is assumed to be at equilibrium before the experiment. At the end of the process, when the ultimate equilibrium state is reached, the gas is distributed homogeneously throughout the complete volume. It is easy to convince oneself that, starting from the latter expanded state, the system does not compress again spontaneously to its initial conditions. This is considered to be the sign of irreversibility. The scenario is a simplified representation of the Joule--Thomson classic experiment (1852) which, however, was not designed for quantitative investigation of the time dependence of the process. By considering only initial and final states, the experiment gives no more than a hint of the existence of a direction to the variable {\em time}.

The purpose of Joule and Thomson was to measure heat produced and exchanged with an external reservoir during spontaneous expansion. For an ideal gas of non--interacting particles, if no mechanical work is allowed to be performed while the system reaches its final equilibrium state, the total exchange of heat with the surroundings will be zero. This fundamental phenomenological result led to the hasty conclusion that irreversible expansion from the initial state to final equilibrium does not involve the surroundings. With the definition of isolation as the condition of a system where neither heat, nor energy under any form (work), nor matter (closed system) is exchanged with the environment, and generalising the conclusion, it has been claimed that irreversible expansion and dispersion of the gas throughout the volume towards final equilibrium is a genuine property of isolated macroscopic systems. From then on, the assignment of a correct mechanism to the process and the justification of its time dependence are considered to be the sole remaining open questions.

In their experiment, Joule and Thomson coupled their system to a reservoir representing the surroundings. The assumption of isolation is therefore incorrect. The presence of this supplementary device allows the exchange of random fluctuations, which may concern energy or momentum. Zero total energy transfer is obtained only by averaging on a time basis much longer than the system's high interaction frequency with the reservoir. If, for historical reasons, it is felt that the word ``isolation'' should still be used in the context described above, it must be qualified by the adjective ``weak''. ``Strict isolation'' should be reserved for objects that are left completely alone.

Let us repeat the experiment under conditions that separate the overall process into its parts. To that end, we examine the effect of rupturing an air--inflated balloon inside either an acoustic reverberation hall or an anechoic chamber. The same picture is obtained by performing the experiment in a room stripped of all curtains, rags or soft tissues on the walls or in the same room, but with the walls covered with soft material. The experiment is no different from Joule's, except that possible intermediate steps of the global dynamics are made observable as the modified fate of acoustic transients.

In the two cases, excess air contained in the balloon disperses irreversibly throughout the rooms. Initial and final conditions are identical in the two cases, as are the nature and physical properties of the gas. The intermediate dynamics appears however to be very different indeed. In the acoustic reverberation hall an acoustic perturbation is created and, the better the walls' reflecting quality, the longer it lasts, limited only by well--known sound absorption. By contrast, in the anechoic room, the perturbation vanishes promptly.

The indisputable experimental fact that the global relaxation dynamics of an expanding gas towards its final equilibrium conditions can be modified by changing the kind of object (curtains versus hard walls) which represents the unavoidable reservoir with which the system interacts, indicates that the process consists of at least two distinct elementary steps. At least one of these depends strongly on the nature of the walls. The weaker the walls' isolating character (soft material), the faster is the global relaxation. The prominent role of the environment is thereby emphasised.

The experiment asserts further that, next to dispersion of the initial perturbation (air compressed in the balloon) throughout the system, full relaxation implies intervention of the walls where correlated acoustic motions must be dissipated. Complete isolation is impossible. In nearly isolated conditions, the second step controls the dynamics but, if the system is strongly coupled with its surroundings, internal redistribution of density and thermal perturbations becomes rate determining. 

The two steps of the global process are very different in their dynamics. Depending on the system that is considered and on the quality of the interaction with the surroundings, they may be almost concomitant. For simplicity, we shall discuss them here as if they were separated in time.

As soon as the membrane separating the two parts of the initial system has been ruptured, a stream of gas is ejected from the higher--pressure compartment, creating a collective and correlated motion of the particles. The system behaves as if it were isolated. If expansion had been performed against a moving piston in adiabatic conditions (no heat exchanged with the environment), work would have been done and therefore energy would have been delivered to the outside world. If the gas were to expand in a vacuum, the same work would have been performed by the system on itself. A jet would have been is created. In the simplified experiment proposed here, where expansion occurs in a low pressure environment, a shock is produced. In both cases, the system provides energy adiabatically for this collective motion. Since the total energy is conserved during this first step (isolated Hamiltonian system, i.e. Newtonian mechanics), that stored now in the collective mode has been taken from the initial thermal supply.

On reaching the wall opposite the rupture, if this is hard, rigid and strictly immobile so that collisions are perfectly elastic, the jet or the shock are reflected and the initial collective motion progressively becomes an acoustic perturbation with the same energy. The spectrum and phases of this motion (coherence) constitute the memory of the initial conditions and of the shape of the reverberating walls.

Depending on the shape of the room where possible acoustic resonance might occur, initial expansion and dispersion is irreversible only in Poincar\'{e}'s sense. This means that the global trajectory does not allow concentration of the particles back into their initial volume for a reasonable period of time. The memory of the initial conditions is however still present in the collective motion, no matter how intricate (chaotic) the individual trajectories may be. The process is apparently irreversible but in fact is not so. Let us call this {\em weak irreversibility}.

Relaxation of the coherent or collective component of the motion starts now. Energy accumulated first in the jet and later in the acoustic perturbation is to be reinjected into the thermal energy bath. The mechanism involves collisions of the system's particles with boundary atoms. Thermal (random) motion of the wall atoms is by no means correlated with that of the system's particles. The trajectories following surface collisions are therefore completely uncoupled with the incoming ones. The events cause stochastic jumps between trajectories. The loss of time correlation near the surface is transmitted to the interior of the system as soon as a particle leaving the surface collides with particles in the bulk.

With ideal gases (hard spheres allowed), loss of coherence thermalises the initial collective motion and returns its energy to the thermal bath. When the collective motion is completely neutralised, the thermal energy of the system (its temperature) has regained the value which it had before expansion, in full agreement with Joule's observation. In the same time, information about the initial conditions is completely lost. This is the sign of {\em strong irreversibility}.

Permanent rapid stochastic intervention of the environment, leading to random transitions between trajectories, blurs the exact conditions of the system in terms of the actual positions and velocities of its constituent particles on a longer time scale. In contradiction with recent literature, the resulting situation is by no means to be interpreted as ``deterministic chaos'', which indeed preserves strictly individual trajectories in the course of time. Instead, the mechanism generates exactly what Boltzmann looked for as ``molecular chaos''. Due to the wealth of different trajectories, all randomly accessible by action of the surroundings, motion can only be discussed in terms of probability distributions of the possible trajectories.

Reference to probability distributions in the context of relaxing objects is not new (Prigogine~{\em et al.} 1988)\nocite{prig:88}. However, the literature refers to uncertain knowledge of initial conditions (at $t = 0$) and not to stochastic mechanical jumps as mentioned above.

\section{Objections} 

The dominant role of the surroundings in the time evolution of macroscopic systems is not readily accepted by the scientific community. Laplace's comment saying {\em ``given precise knowledge of the initial conditions, it should be possible to predict the future of the universe''} remains profoundly rooted in the heart of many theorists. They are not keen on abandoning determinism and, at the same time, losing tight internal control on dynamics. The objections most often cited are listed below together with counter--arguments.

\subsection{Transport coefficients} 

Transport coefficients are major parameters governing the dynamic properties of fluids (gases and liquids). They describe how thermodynamic forces give rise to corresponding flows. Most important are viscosity and thermal conductivity. With multi--component systems out of equilibrium, the diffusion coefficient describes how fast one component flows with respect to the other. Sometimes, flows of different properties are coupled. A typical example is the flow of matter driven by a temperature gradient.

An often expressed objection to our interpretation of Joule's experiment concerns the implication to the transport coefficients. The laws of hydrodynamics predict that sound is damped by viscosity and thermal conductivity, both properties independent of the nature and shape of the fluid's container. Viscosity and other transport effects occurring in the bulk of a fluid are experimentally verified. It is said that their intervention in the phenomenological laws of hydrodynamics (Navier--Stokes equations) does not require the boundaries to be specified. 

The objection treats the transport coefficients as bulk phenomenological parameters, thereby showing confusion between the phenomena and their mechanism. Let us develop this matter by considering the property ``viscosity''. This is the transport coefficient for shear momentum through the fluid (ratio between the rate of transport of shear momentum across the system and gradient of shear velocity, as forced on the system by some unspecified means). As such, it is indeed a bulk property. At the same time, the word more generally expresses viscous flow itself. This implies that what is to be transported is supplied at some places and removed elsewhere. Without the presence of suitable sources and sinks, represented by appropriate boundaries, walls, or analogous interfaces, the very concept is meaningless. In fact, the objection mentioned above is unfair. If it is correct that the differential equation of hydrodynamics do not mention explicitly the presence of boundaries, their solution impies boundary conditions to be specified. 

How is viscosity measured? This may be done, following Couette, by studying the fluid bounded by two parallel plates moving in opposite directions and driven by measurable forces or, following Poiseuille, by measuring the flow driven by a pressure gradient in a capillary. Determination of the value of the coefficient of viscosity is unthinkable without the presence of the boundaries (pair of plates or capillary) on which mechanical force is exerted and measured. Uncoupling viscous flow from the boundaries or walls that are necessary for it to manifest itself is therefore incorrect.

It is sometimes objected that transport coefficients, like viscosity and heat conduction, depend on the collision frequency in the bulk, suggesting that they are by no means related to surface effects. With liquids, the discussion is more complex. Let us therefore consider gaseous systems at moderate pressure, where deviations from the ideal state are negligible. In such systems, the numerical values of the coefficients do depend on the average collision cross section. We must stress that the collision cross--section appears in the denominator of the relevant expressions. Considering that the average collision frequency increases with increasing collision cross section, it is clear that increasing collision frequency reduces transport efficiency, in contradiction with the objection attributing viscocity to collisional effects. 

\subsection{Mixing} 

Mechanical description of macroscopic systems of particles requires a geometrical construction where positions and velocities (better: momenta) of all the particles may be represented and in which the relevant trajectories unfold in the course of time. This construction is called the phase space (not to be confused with configuration space, limited to the position co--ordinates).

Much scientific material has been accumulated in the last two decades on so--called mixing properties of a number of model systems. Starting from simple initial configurations in phase space ($t = 0$), mixing is the property according to which the dynamics would spread the particles progressively as uniformly as possible over the accessible domain. The Sinai billiards and the Lorentz gas are very popular research subjects in this context (Sinai~1976, Spohn~1980, Cornfeld,~{\em et al.}~1982).\nocite {sinai:76,spohn:80,cornfeld:82} They concern computer simulations of parallel beams of non--interacting particles assumed to be moving with fixed velocity in a space containing convex obstacles with which collisions are taken to be elastic (deterministic). Besides the conservation of energy on impact, the dynamics implies conservation of the velocity component parallel to the surface at the collision site (specular reflection). The positions (periodic or random) and the shape of the obstacles determine the reorientation of the particles' velocity at each impact. The convex character of the obstacles results in a complex dispersion of the initial beam in all directions of the configuration space thus destroying the initial collimation.

Many authors stress the importance of mixing on the evolution of representative points in phase space for the restoration and maintenance of chaos or the establishment of ergodic distributions in macroscopic systems. 

It should be stressed, however, that the mathematical definition of mixing involves a phase space, a measure on it, and a group of transformations implying the complete set of dynamical variables. When applied to isolated Hamiltonian systems, published demonstrations never embrace the complete set of canonical variables (positions and momenta) as required by the mathematical definitions. One of the variables of the motion, the magnitude of the velocity, is indeed invariant under the prescribed dynamics. Hence, when applied to dynamics of systems of particles, mixing eludes systematically one of the degrees of freedom in phase space, in contradiction with the fundamental theorems involved. The conclusions are therefore unacceptable.

The mechanical effect of such filters on the motion of particles is pictured exactly by the effect of shining a laser (coherent) beam on diffusing objects. Diffusion spreads the light in all directions but the coherence of the motion is by no means affected (as it would be if it were changed into ``thermal'' light). Strict conservation of temporal coherence can be demonstrated by causing this diffused light to interfere with the incident beam or with another laser beam as in the production of holograms.

Some authors (Balescu~1975, Prigogine {\em et al}~1988)\nocite {balescu:75,prig:88} insist on the mathematical mixing properties of the so--called baker transformation. It is said that, by repeatedly folding a system on itself, as a baker does with dough, initial inhomogeneities are progressively neutralised. In physics, adopting this conclusion is equivalent to cutting off arbitrarily the higher frequency domain of a spectrum because the wavelength would have become too short for the resolving power of the instrument which the observer happens to be using. This is objectively unacceptable since it submits physical reality to subjective implications.

\subsection{Renormalisation} 

Some theoretical trials for justifying the irreversible character of isolated macroscopic dynamics suggest a renormalisation of the system's parameters. The arguments are developed systematically for $N \rightarrow \infty$ and $V \rightarrow \infty$ (Balescu~1975, Goldstein~{\em et al.} ~1975). \nocite{balescu:75, goldstein:75} It is claimed explicitly that {\em this is the only precise way of removing unessential complications due to boundary effects, etc.} (Lanford~1975).\nocite{lanford:75} Infinite systems are regarded as approximations to large finite systems. Such limiting conditions are often labelled ``the thermodynamic limit'', (not to be confused with the same terminology used in hydrodynamics where it characterises a system of which the physical dimensions exceed the mean free path by orders of magnitude).

With Hamiltonian (Newtonian) dynamics, forces are backed by reactions equal in magnitude and opposite in direction. During collisions, the reaction to the force acting on one particle and carried along by its collision partner may be thought of as if shared by the $N-1$ remaining particles. In systems interacting with the surroundings, the reaction is taken care of by dissipation to the boundaries. If we deliberately suppress the role of the boundaries and leave the remaining particles to handle the reaction, which would then be eliminated because its individual incidence on single particles is diluted by their great number, we incorrectly reject an infinite sum of infinitesimal contributions which add up to the value of the reaction force.

\section{The dynamical equation} 

It is now clear that the mechanism by which past events are forgotten as time goes on is related to chaotic evolution. The first question in the debate concerned the kind of chaos implied (deterministic or stochastic). Phenomenological arguments given above, based on variation of the rate of loss of the memory of earlier events, illustrate that deterministic chaos (hypothetically fully isolated systems) cannot lead to the observed irreversible behaviour of nature. However, nothing contradicts the establishment of molecular chaos by stochastic interaction with the fluctuating environment. In this section, we evaluate quantitatively the relaxation dynamics.

From Newton and his followers (Lagrange, Hamilton, etc.) we have learned that the dynamics of particle motions involves their positions and momenta (velocities). This collection of canonical co--ordinates defines the phase space, the points of which represent the complete variety of conditions the system may assume. Starting from some initial set of values of the canonical co--ordinates (at $t = 0$), the equations of the motion describe how these co--ordinates change in the course of time.

With isolated systems, deterministic equations of motion describe the trajectory to be followed by a representative point in phase space. This covers only a portion of phase space in which many trajectories coexist. An elementary theorem of deterministic mechanics states that different trajectories in phase space never cross and that jumps between trajectories are forbidden. In contrast, if systems are allowed to undergo stochastic interactions with the surroundings, these events interrupt existing trajectories and start new ones, with possibly very different conditions. We have stochastic jumps.

Stochastic jumps occur whenever a particle collides with a boundary. In realistic conditions (macroscopic systems) corresponding interactions are so frequent that usual physical measurements on the system, where some time averaging is necessary in order to eliminate fluctuations, easily cover many accessible trajectories. Apart from exceptional cases, single trajectories are indeed extremely short--lived features and are therefore usually irrelevant. In contrast, some of the trajectories available in phase space are more probable than others. The same conclusion holds with individual positions along phase space trajectories (set of co--ordinates). We therefore need to study the time--dependence of probability densities or distribution functions of accessible positions in phase space.

A simple example, that of a particle translating back and forth between walls, should clarify the problem. At time $t = 0$, it is supposed that the velocity of the particle is $v$ (momentum $p = mv$, kinetic energy $E = mv^2/2$). While moving at constant velocity on its initial trajectory, the particle hits the wall. From there, it is reflected but certainly not in the same way as a light beam from a mirror. Depending on the particular motion of the wall atom involved in the collision, the kinetic energy may be modified. It may be increased or decreased. By collision with the wall, the initially sharply peaked probability distribution of energy becomes diffuse, depending on the wall temperature. In fact, the wall and system temperatures equalise.

Not only is the kinetic energy involved. The average direction of the new reflected trajectory also depends on the motion of the wall atoms at impact. While the particle may leave the collision site in any direction, it is expected that, on average, the new trajectory adopts preferentially (highest probability density) a direction corresponding to the average motion of the wall itself.

This simple example highlights the two constituent parts of the equation of the motion, generally valid for all macroscopic systems interacting with their surroundings. At first, in the time separating collisions with the walls, the dynamics strictly follows the laws of deterministic mechanics, with conservation of energy and momentum. The second part concerns impact with the boundary (creating so--called boundary conditions). The former trajectory disappears as in a sink and is replaced by a new one, as if a source were located at the same place. Hence, by writing $f$ for the probability density in phase space, the equation of the motion reads 
\begin{equation} 
\frac{df}{dt}=[f,H]+J. 
\end{equation} 
Symbol $[f, H]$ is the Poisson bracket describing the conservative contribution to the motion
\begin{equation}
[f_N,H] \equiv \sum_r(\frac{\partial H}{\partial p_r}\frac{\partial f_N}{\partial q_r} - \frac{\partial H}{\partial q_r}\frac{\partial f_N}{\partial p_r}),
\label{poisson}
\end{equation}
while $J$ is a source/sink term which explicitly expresses the action of the environment. The equation should be considered to be averaged over the stochastic interaction frequency.

Though difficult to use, mainly because of its generality, this equation allows a very simple discussion of time--dependent phenomena. In stationary conditions $f$ is time independent ($df/dt = 0$). If, on average $J = 0$ (and $[f,H]$ too), the system is at equilibrium and behaves as if it were not interacting with the environment. If the latter conditions are not true, we are in stationary conditions. If $df/dt \neq 0$, we have a transient state. The balance of the two contributions gives the rate of relaxation of the transient. As neither component is negligible, it is difficult to obtain accurate results with this equation. However, in all cases the main conclusion to be drawn from the existence of the sink/source contribution is the prevailing local equilibrium between the system and its surroundings in the boundary regions for all exchangeable properties. This is the analogue (and justification) of the boundary conditions of conventional hydrodynamics.

In hypothetically isolated conditions, $J$ vanishes and the mechanics is purely deterministic and conservative. There is no relaxation.

\section{Thermodynamics} 

In the previous section, the distribution function $f$ was not specified. Accurate conclusions concerning the time dependence of relaxing systems are readily obtained by first studying the shape of $f$ in general.

\subsection{Entropy} 

Referring to earlier work by Carnot and Joule, Clausius invented (1854) the concept of entropy as a special thermodynamic function of state that would characterise macroscopic systems. The change of entropy under modification of a system's conditions was found to depend on the kind of process involved. If this is reversible, we have $\Delta S=Q/T$, where $Q$ is the heat absorbed by the system from the surroundings during the process and $T$ its temperature. By contrast, if the process is irreversible or spontaneous, Clausius observed that $\Delta S>Q/T$. Entropy, as a function of state therefore occupies a key position in discussion of irreversibility in macroscopic dynamics and evolution in the course of time.

According to Clausius, a process is said to be reversible (quasi--stationary) when it is conducted either in such a way that collective or coherent motion is never allowed to develop or when it has been allowed to relax completely (very slow modification of the system's properties). For historical reasons, collective motions are not welcome in traditional equilibrium thermodynamics. The current proposal to generalise thermodynamics to non--equilibrium systems should help reconsideration of this limitation.

In 1877, Boltzmann derived an expression that relates experimental entropy to statistical properties. This reads 
\begin{equation} 
S=k_B \ln W(M), 
\end{equation} 
where $W$ is the measure of the portion of phase space occupied by the particular condition of the system. Another expression for the same symbol is {\em the number of different phase space trajectories complying with the given set of mechanical extensive constraints}. This depends on the nature and the values of the constraints imposed on the system's dynamics. Hence, if we consider a gaseous system in an immobile container of volume V, consisting of $N_r$ particles of type $r$, the total kinetic energy $E$ being exclusively thermal (no additional collective or coherent motion), the set of constraints is the usual collection ($V, N, E$) of microcanonical variables.

\subsubsection{Equilibrium} 

Using for the set of extensive constraints the compact notation $\{X_j\}$, the entropy differentiates as follows 
\begin{equation} 
dS=-\sum_j\xi_j dX_j. 
\end{equation} 
The set of partial differentials $\{\xi_j\}$ is, by definition, the set of intensities or intensive variables conjugate to the microcanonical variables $\{X_j\}$. The distinction between intensive and extensive variables is important.

Rewriting the formal expression and giving to the variables their usual names, Gibbs' traditional equation is retrieved. ($\mu =$ chemical potential). 
\begin{equation} 
dS=\frac{1}{T}dE +\frac{p}{T}dV -\sum_r \frac{\mu_r}{T}dN_r.
\end{equation} 
Equilibrium conditions (no collective motion) having been assumed at the very start, the last expression is an equality.

\subsubsection{Non--equilibrium} 

In non--equilibrium conditions, the complete description of a system's properties requires specification of the additional constraints represented by collective or coherent motion in which part of the total energy is stored. In the example of a spontaneously expanding gas, this was first the jet, followed by the acoustic perturbation. Their presence therefore implies additional extensive properties which must be considered in the expression for the entropy. By differentiation of the entropy with respect to the relevant variables, new conjugate intensities are obtained, typical for the non--equilibrium perturbations.

For simplicity, we consider the particular case of a jet defined by its total momentum ${\mathbf P}$. Gibbs' expression for $dS$ must then be amended by adding a new contribution ($-\xi dX$) which, in this particular case, is $-({\mathbf v}/T) \bullet d{\mathbf P}$, where ${\mathbf v}$ is the average (collective) velocity of the jet (de Hemptinne~1992) \nocite{xh:92}. The momentum in the jet being ${\mathbf P}=N_rm{\mathbf v}$, the additional contribution may also be written $-(1/T)dE_{co}$, where $E_{co}$ is the energy in the collective motion. Hence, in the particular non--equilibrium condition assumed we have: 
\begin{equation} 
dS=\frac{1}{T}dE+\frac{p}{T}dV- \sum\frac{\mu}{T} dN- \frac{1}{T} dE_{co}. 
\end{equation} 
$E$ is now the total energy, the sum of the thermal and the collective contributions. This conclusion is generally valid for all kinds of collective motion.

If the last term had been omitted, if our only information were that the system is definitely in a state of non--equilibrium without our knowing what additional perturbation was relevant, the equal sign would have to be replaced by $\geq$ as Clausius prescribed in his definition of the entropy. This confirms Carnot's statement (Carnot 1827) according to which energy in a collective mode may decrease ($dE_{co} \leq 0$) and thermalise, while the reverse is impossible. \nocite{carnot}

The expression shows that suppression of the collective mode characterising the state of non--equilibrium maximises the entropy. The mechanism involved to reach that maximum is that discussed above, namely decorrelation of the system's internal collective motion by interaction with the surroundings.

\subsection{Distribution function} 

It has been stressed above that a correct definition of the distribution function $f$ requires a thermodynamic debate. This involves maximising the entropy under the given conditions (existence of the collective transient).

Let us consider an arbitrary system of particles at a given instant in unspecified conditions out of equilibrium. The motion of its particles is characterised by many kinds of correlation. In the examples above, stress has been laid only on collective translation. Vortices, internal rotations and vibrations, and a wealth of other motions may contribute to the non--equilibrium conditions. We call the particular state of the system a {\em fluctuation}.

The common tendency of all collective motions is to thermalise. The fundamental mechanism is the same: interaction with a correlation--destroying neighbourhood. The efficiency of the interaction will depend on the kind of motion involved. Some perturbations relax promptly, others last longer. Depending on the time--resolution the observer chooses to consider, which will be in all cases longer that the average interaction frequency with the surroundings, the memory of the fastest components of the initial fluctuation disappears and only the very few longer--lived correlations remain. These are the additional extensive constraints to be considered in maximising the entropy.

Maximisation of the entropy as given by Boltzmann's formula, taking account of all the extensive constraints, including those imposed by collective motion, leads readily to Gibbs' expression for the entropy, based on the distribution function $f$. 
\begin{equation} 
S=k_B \int (f - f \ln f) d \Gamma. 
\end{equation} 
The variables in the maximisation are the set of intensities conjugate to the initially introduced extensive constraints ($d \Gamma$ is the elementary measure in phase space). It is no longer the energy that is important, but the temperature. Energy fluctuates and is therefore known only as an average quantity, while the temperature is defined unambiguously by the surrounding reservoir. The same rule holds for all other constraints strongly coupled to the surroundings.

The function $f$ must now be incorporated in the dynamic equation describing evolution with time of the intensities characterising the state of non--equilibrium the observer has chosen to investigate by selecting an appropriate time resolution. The procedure is straightforward in fluid dynamics where it leads to results in perfect agreement with experiment, both in stationary and in transient conditions (de Hemptinne 1992).\nocite{xh:92}

\section{Quantum mechanics} 

In microphysics (molecular and sub--molecular level), it is known that Newtonian mechanics does not work and must be replaced by quantum mechanics. The most striking property of quantum mechanics is, for many, Heisenberg's uncertainty principle which states that position ($q$) and momentum ($p$) are defined with an uncertainty connected by the relation 
\begin{equation} 
\delta q \; \delta p = h. 
\end{equation} 
For any individual degree of freedom, the space occupied by a single quantum state in phase space equals indeed Planck's constant $h$ (R.K.Pathria 1972 \nocite{pathria:72}). Some authors erroneously replace the equal sign by $\geq$, thereby increasing the impression that quantum mechanics is dominated by uncertainty. This is sometimes taken as the origin of the kind of chaotic uncertainty necessary to justify irreversibility of spontaneous processes.

The formal starting point of quantum mechanics is Schr\"odinger's equation. Only its time--dependent version is relevant here. Let ${\cal H}$ be the Hamiltonian operator. Its structure contains a special contribution for kinetic energy added to the potential field interacting which the particles. Integration leads to a set of functions $\Psi(q,t)$ of position and time (the wave function). The time--dependent Schr\"odinger equation reads 
\begin{equation} 
{\cal H} \Psi (q,t) = - \frac{\hbar}{i} \frac {\partial \Psi (q,t)} {\partial t}. 
\end{equation}

The square of the wave function is usually interpreted as a probability density in configuration space. Mention of probabilities gives a supplementary hint in the direction of fundamental uncertainty, be it only in configuration space.

It can be demonstrated that, for isolated systems, Schr\"odinger's equation is symmetrical with respect to inversion of the variable $t$. This indicates that quantum mechanics alone by no means justifies the irreversible property of nature and that Heisenberg's uncertainty principle refers to another reality.

Schr\"odinger's equation yields stationary states. Transitions between them may be allowed in certain conditions but this implies normally emission or absorption of electromagnetic radiation. If the field, together with a radiating particle, is enclosed in a cavity and supposed to be isolated from the outside world, the solution of Schr\"odinger's equation is a permanent Rabi oscillation back and forth between the previously mentioned stationary states. There is no relaxation (M.~Sargent III {\em et al.}~1974, de~Hemptinne~1985) \nocite{sargent:74, xh:85}.

Decay by emission of radiation, which represents relaxation of excited states, implies that the system would be accessible for exchange of radiation with the outside world. The time--dependent Schr\"odinger equation confirms that the outgoing field is phase--matched to the motion of the radiating particles (coherence). At the same time, in addition to the ubiquitous background radiation, the field accessible for re--absorption is the total incoherent thermal radiation issued from the collection of external sources which constitute the surroundings. As in the classical case, one result of exchange of radiation with the outside world is complete loss of memory of the initial phase information. At equilibrium, the temperature of the set of radiators equals that of the surroundings (de~Hemptinne~1991) \nocite{xh:91}.

In fact, quantum mechanics bridges the motion of particles and the properties of radiation fields. It appears that incoherence of thermal sources of radiation is the analogue of molecular chaos of particle motion.

\section{Conclusions} 

Irreversibility, time's arrow and its origin remain the source of much scientific discussion. This started before Boltzmann. The dispute has been nourished by inaccurate definitions and hasty conclusions concerning a number of concepts such as the word irreversibility itself, isolation, relaxation and equilibrium. The controversy has been amplified by ambiguous interpretations of simple experimental facts (de Hemptinne 1997, Kumi{\v c}{\' a}k {\em et al.} 1998). \nocite{xh:97,xh:98}

Arguments based on observation and theoretical discussions strictly compatible with confirmed laws of mechanics (at microscopic level) have been used in this work to emphasise that the equilibrium state cannot be defined without taking account of the conditions valid in the environment. Relaxation dynamics represents exchange with the surroundings, that is export of coherent (collective) information compensated by stochastic import of thermalised (incoherent) information.

Simultaneous discussion of the properties of the system and its surroundings makes the use of thermodynamic arguments necessary. Intensities, which are directly related to the probabilistic concept of entropy, are defined for every exchangeable property.

It is frequently objected that interaction of the walls and the enclosed system itself follows Hamiltonian dynamics. It would therefore suffice to define the macroscopic system as the addition of the enclosed system and its walls in order to build a (strictly) isolated system where the laws of Hamiltonian dynamics would be strictly valid. The discussion concerning irreversibility would then occur at this higher level, where the global system would be isolated. This is not the case, because the walls themselves interact with a broader environment, moving the problem one step further.

There is an arbitrary choice in the definition of boundaries to systems. It depends on how far mechanics is allowed to take care of correlated reactions to forces. In the domain where Hamiltonian dynamics is valid, forces acting on the components of the system sum identically to zero. Some forces, however, clearly have an external source. Their reaction cannot be included in the dynamic equation unless they are labelled ``stochastic'' because they are completely uncorrelated. They are the cause of irreversibility.

We are tempted to extrapolate questions and conclusions to the universe itself, assuming certain alleged cosmological properties, often without proof. This must be strongly resisted. Science has no information on the extreme properties of the universe and conclusions based simply on the extrapolation of arguments valid at our observational level to domains beyond our reach are meaningless and invalid. This is why the final source of irreversibility will never be unveiled.

\end{document}